# Social Diffusion and Global Drift in Adaptive Social Networks

HIROKI SAYAMA, Collective Dynamics of Complex Systems Research Group, Binghamton University, State University of New York / Center for Complex Network Research, Northeastern University

## 1. INTRODUCTION

Social contagion has been studied in various contexts. Many instances of social contagion can be modeled as an infection process where a specific state (adoption of product, fad, knowledge, behavior, etc.) spreads from individual to individual through links between them [Van den Bulte and Stremersch 2004; Dodds and Watts 2004; Hill et al. 2010]. In the meantime, other forms of social contagion may better be understood as a diffusion process where the state of an individual tends to assimilate with the social norm (i.e., local average state) within his/her neighborhood [Christakis and Fowler 2007; Fowler and Christakis 2008; Coronges et al. 2011].

Unlike infection scenarios where influence is nonlinear, unidirectional, fast, and potentially disruptive, social diffusion is linear, bidirectional, gradual, and converging. The distance between an individual's state and his/her neighbors' average state always decreases, and thus a homogeneous global state is guaranteed to be the network's stable equilibrium state in the long run. This does not sound as intriguing or exciting as infection dynamics, which might be why there are very few studies on mathematical models of social diffusion processes.

Here, this study attempts to shed new light on an unrecognized characteristic of social diffusion, i.e., non-trivial drift it can cause to the network's global average state. Although somewhat counterintuitive, such global drift is indeed possible because, unlike physical diffusion processes, social diffusion processes are *not* conservational. In what follows, a mathematical model of social diffusion will be presented to explain the mechanism of this phenomenon, and some possible collective actions for influencing the direction of global drift will be proposed. The relevance of social diffusion to individual and collective improvement will be discussed briefly, with an emphasis on educational applications.

## 2. MATHEMATICAL MODEL

Let us begin with the traditional diffusion equation on a network,

$$\frac{ds}{dt} = -cLs, \tag{1}$$

where $s$ is the node state vector of the network, $c$ the diffusion constant, and $L$ the Laplacian matrix of the network. If the network is connected, the coefficient matrix $(-cL)$ has one and only one dominant eigenvalue, 0, whose corresponding eigenvector is the homogeneity vector $h = (111\ldots 1)^T$. Therefore, the solutions of this equation always converge to a homogeneous state regardless of initial conditions if the network is connected. Moreover, it can be shown that the sum (or, equivalently, average) of node states, $h^T s$, will not change over time, because

$$\frac{d(h^T s)}{dt} = -ch^T Ls = -c(Lh)^T s = 0. \tag{2}$$

This indicates that no drift of the global average state is possible in simple diffusion processes on a network.

However, social diffusion processes can be modeled slightly differently. Adopting a typical assumption that each individual assimilates his/her state with the social norm around him/her, the dynamical equation of social diffusion can be written individually as

$$\frac{ds_i}{dt} = c \left( \frac{\sum_{j \in N_i} s_j}{k_i} - s_i \right), \tag{3}$$

where $s_i$ is the state of individual $i$, $N_i$ the neighbor set of individual $i$, and $k_i$ the number of $i$'s neighbors (= degree of node $i$). The first term inside the parentheses represents a local average state around individual $i$ that works as the social norm for him/her. This equation can be rewritten at a collective level as

$$\frac{ds}{dt} = -cD^{-1}Ls, \tag{4}$$

where $D^{-1}$ is the inverse degree matrix whose $i$-th diagonal component is $1/k_i$ while non-diagonal components are all zero. This equation is essentially the same as Eq. (1) if the network is regular (i.e., $k_i = k \; \forall i$). Even if not regular, a connected network still has the same stable homogeneous equilibrium state, $h$, as in the traditional diffusion equation. However, the sum of node states, $h^T s$, is no longer conserved in this model:

$$\frac{d(h^T s)}{dt} = -ch^T D^{-1} Ls \tag{5}$$

This is not equal to zero for most $s$ and $D$, which means that the global average state can and do drift in social diffusion processes.

The direction of global drift can be studied using the equation above. With $v = -LD^{-1}h$, Eq. (5) can be written as

$$\frac{d(h^T s)}{dt} = cv^T s, \tag{6}$$

which shows that correlation between two vectors, $v$ and $s$, will determine the direction of global drift. Vector $v$ is further detailed as

$$v = -LD^{-1}h = (A-D)D^{-1}h = (AD^{-1} - I)h = \begin{pmatrix} -1 + \sum_{j \in N_1} k_j^{-1} \\ -1 + \sum_{j \in N_2} k_j^{-1} \\ \vdots \\ -1 + \sum_{j \in N_n} k_j^{-1} \end{pmatrix} = \begin{pmatrix} \langle k_1/k_j \rangle_{j \in N_1} - 1 \\ \langle k_2/k_j \rangle_{j \in N_2} - 1 \\ \vdots \\ \langle k_n/k_j \rangle_{j \in N_n} - 1 \end{pmatrix}, \tag{7}$$

where $A$ is the adjacency matrix, $D$ the degree matrix, $n$ the number of individuals, and $\langle \cdots \rangle_{j \in N_i}$ a local average within neighborhood $N_i$. Each component of $v$ is the difference between the local average of self/neighbor degree ratios and 1, which tends to be positive if the individual has more connections than its neighbors, or negative otherwise. In this regard, vector $v$ characterizes the local degree differences for all individuals in society. If the current state distribution is positively (or negatively) correlated with this vector, the global average state will drift upward (or downward) due to social diffusion.

## 3. COLLECTIVE ACTIONS FOR GLOBAL UPDRIFT

The theoretical conclusion introduced above suggests that, if the numbers of connections people have are manipulated to positively correlate with their states, then the global state will collectively move upward. Such correlation may be artificially introduced in a relatively simple adaptive network [Gross and Sayama 2009] method, e.g., by preferential link reassignment according to node states.

The effect of this adaptive network approach was tested through computational simulations. The simulation model assumes $n$ nodes connected by $m$ undirected links. Their states and connections are initially random. In each iteration, node $i$'s state $s_i$ is updated by a discrete-time version of Eq. (3):

$$s_i(t+1) = s_i(t) + c \left( \frac{\sum_{j \in N_i(t)} s_j(t)}{k_i(t)} - s_i(t) \right) \tag{8}$$

To make the nodes adaptively change their connections, $pm$ links are randomly chosen and removed from the network in each iteration and replaced by the same number of new links between nodes that are selected preferentially based on their states, with selection probability $P(i;t) \sim s_i(t)^\alpha$. Systematic simulations were conducted with $p$ and $\alpha$ varied. Results (Fig. 1) show that the preferential link reassignment with positive preferential selection ($p > 0$, $\alpha > 0$) induced upward drift of the global average state over time (Fig. 1C). Moreover, this upward drift continues indefinitely if small random fluctuations are added to individuals' states (Fig. 1D). This is because adaptive link reassignment tends to preferentially favor nodes that had upward fluctuations by assigning more connections to them.

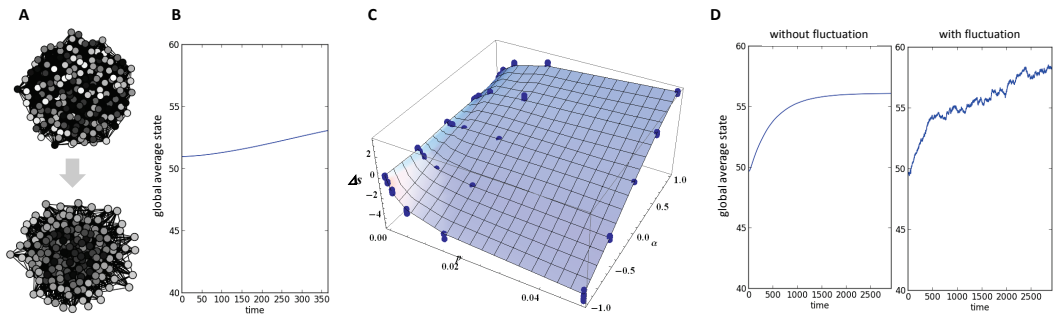

Fig. 1. Computer simulation results with $n = 200$, $m = 4000$, $c = 0.001$. **A.** An example of network evolution with $p = 0.05$ and $\alpha = 1.0$. The network is initially random in both topology and state distribution (top), but over time, a central cluster of nodes with greater state values forms (bottom) while states diffuse. The shade in each node represents the individual's state. **B.** Temporal change of the global average state taken from the simulation run shown in A. **C.** Summary of systematic simulations with varying $p$ and $\alpha$. The difference in average states between before and after a certain number of iterations ($\Delta s$) is plotted over the $p$-$\alpha$ parameter space. **D.** Long-term behavior of global average states in two different conditions. Left: Without random fluctuation of node states. Right: With random fluctuation of node states.

## 4. IMPLICATIONS FOR INDIVIDUAL AND COLLECTIVE IMPROVEMENT

This study has illustrated the possibility that social diffusion may be exploited for individual and collective improvement. Mechanisms like the adaptive link reassignment used in the simulations above may be utilized in practice to, for example, help spread desirable behaviors and/or suppress undesirable behaviors among youths. One particularly interesting application area is education. The author and his collaborators have studied possible diffusion of academic success in high school students' social network [Blansky et al. 2013], which naturally led educators to ask how one could utilize such diffusion dynamics to improve the students' success at a whole school level. Naïve mixing of students may not be a good strategy due to bidirectional effects of social diffusion. Inducing correlation between degrees and academic achievements, on the other hand, may be a practically implementable solution at school by allowing higher-achieving students to participate in more extracurricular activities.